\documentclass[12pt,preprint]{aastex}

\slugcomment{}

\shorttitle{Clathration of Volatiles in the Solar Nebula and Implications for the Origin of Titan's atmosphere}

\shortauthors{Mousis et al.}

\begin{document}

\title{Clathration of Volatiles in the Solar Nebula and Implications for the Origin of Titan's atmosphere}

\author{
Olivier~Mousis\altaffilmark{1},
Jonathan I.~Lunine\altaffilmark{2},
Caroline~Thomas\altaffilmark{1},
Matthew~Pasek\altaffilmark{2},
Ulysse~Marb\oe uf\altaffilmark{1},
Yann~Alibert\altaffilmark{1},
Vincent~Ballenegger\altaffilmark{1},
Daniel~Cordier\altaffilmark{3,4},
Yves~Ellinger\altaffilmark{5},
Fran{\c c}oise~Pauzat\altaffilmark{5}
\& Sylvain~Picaud\altaffilmark{1}}

\altaffiltext{1}{Universit{\'e} de Franche-Comt{\'e}, Institut UTINAM, CNRS/INSU, UMR 6213, 25030 Besan\c{c}on Cedex, France}

\email{olivier.mousis@obs-besancon.fr}

\altaffiltext{2}{Lunar and Planetary Laboratory, University of Arizona, Tucson, AZ, USA}

\altaffiltext{3}{Institut de Physique de Rennes, CNRS, UMR 6251, Universit{\'e} de Rennes 1, Campus de Beaulieu, 35042 Rennes, France}

\altaffiltext{4}{Ecole Nationale Sup{\'e}rieure de Chimie de Rennes, Campus de Beaulieu, 35700 Rennes, France}

\altaffiltext{5}{Universit{\'e} Pierre et Marie Curie, Laboratoire de Chimie Th{\'e}orique, CNRS/INSU, UMR 7616, 75252 Paris Cedex 05, France}

\begin{abstract}
{We describe a scenario of Titan's formation matching the constraints imposed by its current atmospheric composition. Assuming that the abundances of all elements, including oxygen, are solar in the outer nebula, we show that the icy planetesimals were agglomerated in the feeding zone of Saturn from a mixture of clathrates with multiple guest species, so-called stochiometric hydrates such as ammonia hydrate, and pure condensates. We also use a statistical thermodynamic approach to constrain the composition of multiple guest clathrates formed in the solar nebula. We then infer that krypton and xenon, that are expected to condense in the 20--30 K temperature range in the solar nebula, are trapped in clathrates at higher temperatures than 50 K. Once formed, these ices either were accreted by Saturn or remained embedded in its surrounding subnebula until they found their way into the regular satellites growing around Saturn.  In order to explain the carbon monoxide and primordial argon deficiencies of Titan's atmosphere, we suggest that the satellite was formed from icy planetesimals initially produced in the solar nebula and that were partially devolatilized at a temperature not exceeding $\sim$ 50 K during their migration within Saturn's subnebula. The observed deficiencies of Titan's atmosphere in krypton and xenon could result from other processes that may have occurred both prior or after the completion of Titan. Thus, krypton and xenon may have been sequestrated in the form of XH$_3^+$ complexes in the solar nebula gas phase, causing the formation of noble gas-poor planetesimals ultimately accreted by Titan. Alternatively, krypton and xenon may have also been trapped  efficiently in clathrates located on the satellite's surface or in its atmospheric haze. We finally discuss the subsequent observations that would allow to determine which of these processes is the most likely.}
\end{abstract}


\section{Introduction}
\label{sec:intro}

The exploration of Saturn's satellite system by the {\it Cassini-Huygens} spacecraft has forced a reappraisal of existing models for the origin of Titan. Indeed, a puzzling feature of the atmosphere of Titan is that no primordial noble gases other than argon were detected by the Gas Chromatograph Mass Spectrometer (GCMS) aboard the {\it Huygens} probe during its descent to Titan's surface on January 14, 2005. The observed argon includes primordial ${}^{36}$Ar (the main isotope) and the radiogenic isotope ${}^{40}$Ar, which is a decay product of ${}^{40}$K (Niemann et al. 2005).  In any case, with a ${}^{36}$Ar/${}^{14}$N lower than the solar value by more than five orders of magnitude (Niemann et al. 2005), a very tight constraint is imposed on the origin of the bulk constituent of Titan's atmosphere, N$_2$ (Owen 1982). The other primordial noble gases Kr and Xe (and  $^{38}$Ar) were not detected by the GCMS instrument down to upper limits of 10 parts per billion relative to nitrogen (Niemann et al. 2005). These latter nondetections are striking given the detection of noble gases in the atmospheres of telluric planets, as well as in the atmosphere of Jupiter where their abundances are found to be over-solar (Owen et al. 1999). Moreover, the measurements made by the Gas Chromatograph Mass Spectrometer (GCMS) aboard the {\it Huygens} probe have confirmed what was previously known from Voyager and ground-based observations: that the atmosphere of Titan is dominated by N$_2$ and CH$_4$, with a low CO:CH$_4$ ratio (CO:CH$_4$ $\sim$10$^{-3}$; Gautier \& Raulin 1997). This last value is also a strong constraint on the composition because CO is believed to have been more abundant than CH$_4$ in the solar nebula (Prinn \& Fegley 1981), a result indirectly confirmed by studies of the densities of Kuiper Belt and former Kuiper Belt objects (Johnson \& Lunine 2005).

It has been proposed that Titan's atmospheric molecular nitrogen results from ammonia photolysis or shock chemistry early in Titan's history (Atreya et al. 1978; McKay et al. 1988), implying that this species is probably not primordial. In addition, the high CH$_3$D:CH$_4$ ratio measured in the satellite's atmosphere (D:H = $1.32^{+0.15}_{-0.11} \times 10^{-4}$; B{\'e}zard et al. 2007) implies that this compound is likely to have been incorporated in solids produced in the solar nebula prior to having been accreted by Titan. Thus, CH$_3$D originated from interstellar methane highly D-enriched ices that vaporized when entering the solar nebula. Deuterated methane then isotopically exchanged with molecular hydrogen in the gas phase prior to its trapping in solids formed in the nebula (Mousis et al. 2002). The high deuterium enrichment acquired by methane in the satellite's atmosphere excludes its production from the gas phase conversion of carbon monoxide in a dense and warm subnebula, as was first suggested by Prinn \& Fegley (1981). Indeed, in this case the deuterium enrichment of methane produced from carbon monoxide in Saturn's subnebula would be almost solar (Mousis et al. 2002). It would then be impossible to explain the measured value from atmospheric photochemistry only (Cordier et al. 2008). Note that it has alternatively been suggested that the atmospheric methane of Titan could be produced from serpentinization in its interior (Atreya et al. 2006), but there is as yet no isotopic test of this process that has been proposed. Identification of abundant carbon dioxide in fresh volcanic deposits might favor production of hydrocarbons via serpentinization in the interior, and some evidence for such deposits has been argued from Cassini VIMS data (Hayne et al. 2008), but the extent of such deposits remains to be quantified. 

Recently, in order to explain the deficiencies of CO, N$_2$, Ar and Kr in Titan, Alibert \& Mousis (2007; hereafter AM07) proposed that the satellite formed in a Saturn's subnebula warm enough to partly devolatilize the planetesimals captured from the solar nebula itself. Indeed, following AM07, the migration of planetesimals in a balmy subnebula would allow an efficient devolatilization of most volatile species (CO, N$_2$, Ar and Kr) whose clathration temperatures are low, whereas H$_2$O, NH$_3$, H$_2$S, CO$_2$, Xe and CH$_4$ would remain incorporated in solids. However, since Xe was expected to remain trapped in planetesimals, the scenario proposed by AM07 could not explain in a self-consistent way the observed deficiency of this noble gas in Titan's atmosphere.

Moreover, because AM07 did not consider the possibility of multiple guest trapping when they determined the clathration sequence in the solar nebula, their predictions about the incorporation conditions of the different volatiles could be strongly modified by this effect. Indeed, it has been shown in the literature (Lunine \& Stevenson 1985; Osegovic \& Max 2005; Thomas et al. 2007,2008) that minor compounds can deeply affect the composition of clathrates formed from a given gas phase. As a result, the validity of the Titan's formation scenario proposed by AM07 can be questioned if the multiple guest trapping of volatiles is considered during the clathration sequence in the solar nebula.
 
In addition, AM07 made the ad hoc hypothesis that the oxygen abundance was ``oversolar'' in the feeding zone of Saturn in order to provide enough water to allow the full clathration of volatiles during the cooling of the nebula. This assumption was supported by the work of Hersant et al. (2001) who estimated that Jupiter was formed at temperatures higher than $\sim$ 40--50 K in the solar nebula.
Hence, because hydrates and clathrates generally form at higher temperatures than pure condensates, the accretion of these ices was thus required during the formation of Jupiter in order to explain the volatile enrichments observed in its atmosphere (Gautier et al. 2001a,b). The fact that both Jupiter and Saturn might have accreted planetesimals formed at similar locations during their growth and migration within the solar nebula (Alibert et al. 2005a) thus led AM07 to consider the same gas phase composition in the two giant planets feeding zones. However, Hersant et al. (2001) only used an evolutionary solar nebula model to derive the disk's temperature at the time when the mass of Jupiter's feeding zone was equal to that of the gas in its current envelope, and did not include the influence of protoplanet formation on the structure of the disk. Recent giant planet core accretion formation models that include migration, disk evolution, such as those proposed by Alibert et al. (2004,2005a), have shown that the nebula's temperature can be as low as $\sim$ 10--20K at the end of Jupiter's formation. This implies that, during their formation, both Jupiter and Saturn can accrete pure condensates produced at lower temperatures than those required for clathration. As a result, no extra water is required in the nebula to allow all the volatiles to be trapped in clathrates, and the over-solar oxygen abundance condition in the nebula can be relaxed.\\

In this paper, we reinvestigate the scenario of Titan's formation initially proposed by AM07 in order to match the constraints derived from its atmospheric composition. We assume that the abundance of all elements, including oxygen, are solar in the outer nebula, implying that the icy planetesimals were agglomerated in the feeding zone of Saturn from a mixture of clathrates with multiple guest species, hydrates such as ammonia hydrate, and pure condensates. In addition, we use a statistical thermodynamic approach to constrain the composition of clathrates formed in the solar nebula. We then infer that krypton and xenon, that are expected to condense in the 20--30 K temperature range in the solar nebula, are trapped in clathrates at higher temperatures than 50 K. Once formed, these ices either were accreted by Saturn or remained embedded in Saturn's surrounding subnebula until they found their way into the regular satellites growing around Saturn.  In order to explain the carbon monoxide and primordial argon deficiencies of Titan's atmosphere, we suggest that the satellite was formed from icy planetesimals initially produced in the solar nebula and that were partially devolatilized at a temperature not exceeding $\sim$ 50 K during their migration within Saturn's subnebula. On the other hand, the observed deficiencies of Titan's atmosphere in krypton and xenon must result from other processes that  occurred both prior or after the completion of Titan. \\

We determine the formation sequence of the different ices produced in the outer solar nebula and then examine the relative propensities among the different volatiles for trapping in clathrates
produced in Saturn's feeding zone in \S~\ref{sec:ices}, assuming solar abundances for all elements. In \S~\ref{sec:Titan}, we propose that carbon monoxide and argon only were devolatilized from planetesimals, due to the increasing gas temperature and pressure conditions they  encountered during their migration inwards in the Saturn's subnebula. We then explore the different mechanisms that may explain the deficiency of Titan's atmosphere in krypton and xenon. \S~\ref{sec:discussion} is devoted to discussion.

\section{Formation of icy planetesimals in Saturn's feeding zone}
\label{sec:ices}

In this Section, we first describe the formation sequence of the different ices produced in the outer solar nebula. We then examine the relative propensities among the different volatiles for trapping in clathrates produced in Saturn's feeding zone at temperatures greater than $\sim$ 50 K. Once formed, these ices will add to the composition of the planetesimals that will be either accreted by Saturn or embedded in its surrounding subnebula from which the regular satellite system has formed.

\subsection{Initial gas phase composition}
\label{sec:gas}

In order to define the gas phase composition in the feeding zone of  Saturn, we assume that the abundances of all elements are solar (Lodders 2003; see Table \ref{lodders}) and consider both refractory and volatile components. Refractory components include rocks and organics. According to Lodders (2003), rocks contain $\sim$ 23\% of the total oxygen in the nebula. The fractional abundance of organic carbon is assumed to be 55\% of total carbon (Pollack et al.\ 1994), and the ratio of C:O:N included in organics is supposed to be 1:0.5:0.12 (Jessberger et al.\ 1988). We then assume that the remaining O, C, and N exist only under the form of H$_2$O, CO, CO$_2$, CH$_3$OH, CH$_4$, N$_2$, and NH$_3$.  Hence, once the gas phase abundances of elements are defined, the abundances of CO, CO$_2$,  CH$_3$OH, CH$_4$, N$_2$ and NH$_3$ are determined from the adopted CO:CO$_2$:CH$_3$OH:CH$_4$ and N$_2$:NH$_3$ gas phase molecular ratios, and from the C:O:N relative abundances set in organics. Finally, once the abundances of these molecules are fixed, the remaining O gives the abundance of H$_2$O.

We then set CO:CO$_2$:CH$_3$OH:CH$_4$ = 70:10:2:1 in the gas phase of the disk, values that are consistent with the ISM measurements considering the contributions of both gas and solid phases in the lines of sight (Frerking et al. 1982; Ohishi et al. 1992; Ehrenfreund  \& Schutte 2000; Gibb et al. 2000). In addition, S is assumed to exist in the form of H$_2$S, with H$_2$S:H$_2$ = 0.5 $\times$ (S:H$_2$)$_\sun$, and other refractory sulfide components  (Pasek et al. 2005). We also consider N$_2$:NH$_3$ = 1:1 in the nebula gas-phase. This value is compatible with thermochemical calculations in the solar nebula that take into account catalytic effects of Fe grains on the kinetics of N$_2$ to NH$_3$ conversion (Fegley 2000).

\subsection{Formation of ices in Saturn's feeding zone}

The process by which volatiles are trapped in icy planetesimals, illustrated in Figs. 1 and 2, is calculated using the stability curves of hydrates, clathrates and pure condensates, and the thermodynamic path detailing the evolution of temperature and pressure at 9.5 AU in the solar nebula, corresponding to the actual position of Saturn. This thermodynamic path (hereafter cooling curve) results from the determination of the disk vertical structure which, in turn, derives from the turbulent model used in this work.  While a full description of our turbulent model of the accretion disk can be found in Papaloizou \& Terquem (1999) and Alibert et al. (2005b), we simply provide here a concise outline of the underlying assumptions.

Assuming cylindrical symmetry in the protoplanetary disk, the vertical structure along the $z$-axis at a given radial distance $r$ is calculated by solving a system of three equations, namely, the equation for hydrostatic equilibrium, the energy conservation, and the diffusion equation for the radiative flux (Eqs.~(1), (2), and (3) in Alibert et al.\ 2005b, respectively). The system variables are the local pressure $P(r,z)$, temperature $T(r,z)$, density $\rho(r,z)$, and viscosity $\nu(r,z)$. Following Shakura \& Sunyaev (1973), the turbulent viscosity $\nu$ is expressed in terms of Keplerian rotation frequency $\Omega$ and sound velocity $C_s$, as $\nu = \alpha C_s^2 / \Omega$, where $\alpha$ is a parameter characterizing the turbulence in the disk. For the model we use here, the $\alpha$ parameter is equal to $2 \times 10^{-3}$, and the total evaporation rate is of the order of 10$^{-8}$ M$_\odot$/year. At the beginning of the calculation, the gas surface density is given by a power law, $\Sigma \propto r^{-3/2}$, normalized to have $\Sigma = 600$g/cm$^2$ at the current day position of Jupiter. This disk model was used in Alibert et al. (2005a) in order to calculate formation models of Jupiter and Saturn.

The stability curves of hydrates and clathrates derive from Lunine \& Stevenson (1985)'s compilation of published experimental work, in which data are available at relatively low temperatures and pressures. On the other hand, the stability curves of pure condensates used in our calculations derive from the compilation of laboratory data given in the CRC Handbook of chemistry and physics (Lide 2002). The cooling curve intercepts the stability curves of the different ices at particular temperatures and pressures. For each ice considered, the domain of stability is the region located below its corresponding stability curve. The clathration process stops when no more crystalline water ice is available to trap the volatile species. Note that, in the pressure conditions of the solar nebula, CO$_2$ is the only species that crystallizes at a higher temperature than its associated clathrate. We then assume that solid CO$_2$ is the only existing condensed form of CO$_2$ in this environment. In addition, we have considered only the formation of pure ice of CH$_3$OH in our calculations since, to our best knowledge, no experimental data concerning the stability curve of its associated clathrate have been reported in the literature. As a result of the assumed solar gas phase abundance for oxygen, ices formed in the outer solar nebula are composed of a mix of clathrates, hydrates and pure condensates which are, except for CO$_2$ and CH$_3$OH, produced at temperatures ranging between 20 and 50 K. Once formed, the different ices agglomerated and incorporated into the growing planetesimals.

\begin{figure}
\resizebox{\hsize}{!}{\includegraphics[angle=0]{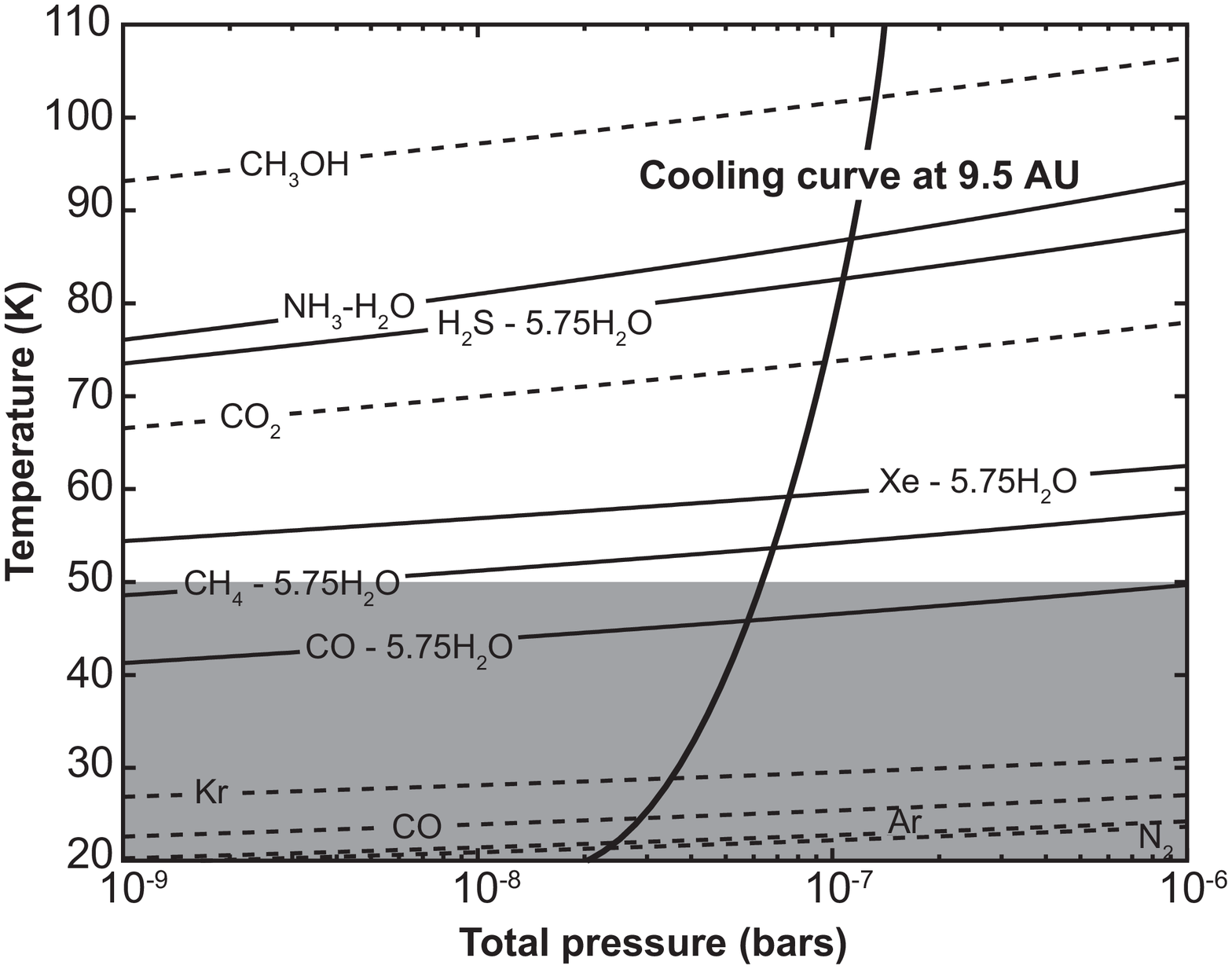}} \caption{Stability curves of hydrate (NH$_3$-H$_2$O), clathrates (X-5.75H$_2$O) (solid lines), and pure condensates (dotted lines), and cooling curve of the solar nebula at the heliocentric distance of 9.5 AU, assuming a full efficiency of clathration. Species remain in the gas phase above the stability curves. Below, they are trapped as clathrates or simply condense. The grey area characterises the different ices assumed to be heated during their migration and accretion in Saturn's subnebula to form proto-Titan.} 
\label{cooling}
\end{figure}

Because the efficiency of clathration remains unknown in the solar nebula, we explore here two opposite cases of trapping efficiencies of volatiles. Figure \ref{cooling} illustrates the case where the efficiency of clathration is total, implying that guest molecules had the time to diffuse through porous water-ice planetesimals in the solar nebula. This situation is plausible if we consider that collisions between planetesimals have exposed essentially all the ice to the gas over time scales shorter or equal to planetesimals lifetimes in the nebula (Lunine \& Stevenson 1985). In this case, NH$_3$, H$_2$S, Xe, CH$_4$ and $\sim$ 60\% of CO form NH$_3$-H$_2$O hydrate and H$_2$S-5.75H$_2$O, Xe-5.75H$_2$O, CH$_4$-5.75H$_2$O and CO-5.75H$_2$O clathrates with the available water. The remaining CO, as well as N$_2$, Kr, and Ar, whose clathration normally occurs at lower temperatures, remain in the gas phase until the nebula cools enough to allow the formation of pure condensates. 

\begin{figure}
\resizebox{\hsize}{!}{\includegraphics[angle=0]{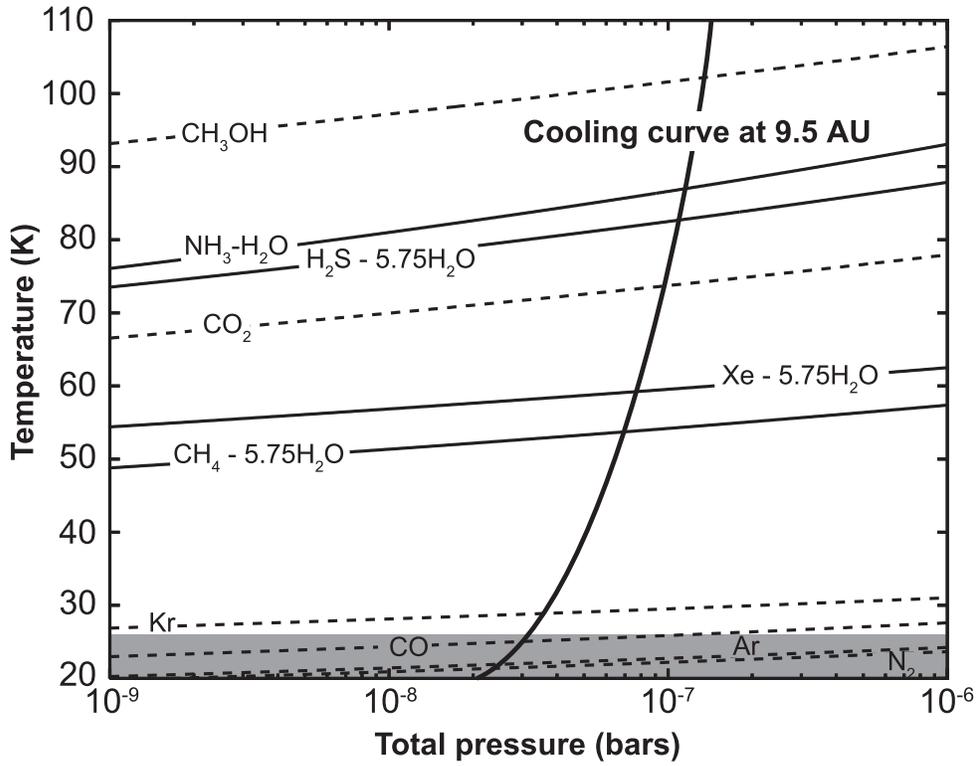}} \caption{Same as Fig. \ref{cooling}, but with a clathration efficiency of $\sim$ 25\%.} 
\label{cooling2}
\end{figure}

Figure \ref{cooling2} illustrates the case where the efficiency of clathration is only of $\sim$ 25\%. Here, either only a part of the clathrates cages have been filled by guest molecules, either the diffusion of clathrated layers through the planetesimals was to slow to enclathrate most of the ice, or the poor trapping efficiency was the combination of these two processes. In this case, only NH$_3$, H$_2$S, Xe and CH$_4$ form NH$_3$-H$_2$O hydrate and H$_2$S-5.75H$_2$O, Xe-5.75H$_2$O and CH$_4$-5.75H$_2$O clathrates. Due to the deficiency in accessible water in icy planetesimals, all CO, Kr and N$_2$ form pure condensates in the solar nebula.

In both cases, i.e. high and poor clathration efficiencies, the abundances of volatiles observed in the envelopes of Jupiter and Saturn can be reproduced in a way consistent with their internal structure models, provided that the formation of the two planets ended at $\sim$ 20 K in the solar nebula (Mousis \& Marboeuf 2006)\footnote{If Jupiter and Saturn were formed at $T \ge $ 50 K in the solar nebula, only hydrates and clathrates, and not pure condensates, would be accreted in their interiors. This assumption has been used by Hersant et al. (2008) to predict an oversolar abundance of Xe in the envelope of Saturn, and solar abundances for Ar and Kr.}. As discussed in \S~\ref{sec:devol}, these two formation sequences of the different ices in the feeding zone of Saturn are compatible with the scenario of partial devolatilization of planetesimals accreted by Titan proposed by AM07.

\subsection{Multiple guest trapping in clathrates} 
\label{MG}

We calculate here the relative abundances of guests that can be incorporated in H$_2$S, Xe and CH$_4$ dominated clathrates at the time of their formation at temperatures greater than $\sim$ 50 K in the solar nebula. In our calculations, any volatile already trapped or condensed at a higher temperature than that of the considered clathrate is excluded from the coexisting gas phase composition. We follow the method described by Lunine \& Stevenson (1985) and Thomas et al. (2007, 2008) which uses classical statistical mechanics to relate the macroscopic thermodynamic properties of clathrates to the molecular structure and interaction energies. It is based on the original ideas of van der Waals \& Platteeuw (1959) for clathrate formation, which assume that trapping of guest molecules into cages corresponds to the three-dimensional generalization of ideal localized adsorption. In this model, the occupancy fraction of a guest molecule $K$ for a given type $t$ of cage ($t$~=~small or large) can be written as

\begin{equation}
\label{occupation}
y_{K,t}=\frac{C_{K,t}P_K}{1+\sum_{J}C_{J,t}P_J} ,
\end{equation}

\noindent where the sum in the denominator includes all the species which are present in the initial gas phase. $C_{K,t}$ is the Langmuir constant of species $K$ in the cage of type $t$, and $P_K$ is the partial pressure of species $K$. This partial pressure is given by $P_K=x_K\times P$, with $x_K$ the molar fraction of species $K$ in the initial gas phase given in Table \ref{lodders}, and $P$ the total H$_2$ pressure. The Langmuir constants are in turn determined by integrating the molecular potential within the cavity as

\begin{equation}
\label{langmuir}
C_{K,t}=\frac{4\pi}{k_B
T}\int_{0}^{R_c}\exp\Big(-\frac{w(r)}{k_B T}\Big)r^2dr ,
\end{equation}

\noindent where $R_c$ represents the radius of the cavity assumed to be spherical, and $w(r)$ is the spherically averaged Kihara potential representing the interactions between the guest molecules and the H$_2$O molecules forming the surrounding cage. Following McKoy \& Sinano\u glu (1963), this potential $w(r)$ can be written for a spherical guest molecule, as

\begin{equation}
\label{pot_Kihara}
w(r) = 2z\epsilon\frac{\sigma^{12}}{R_c^{11}r}\Big(\delta^{10}(r)+\frac{a}{R_c}\delta^{11}(r)\Big)\\
 - \frac{\sigma^6}{R_c^5r}\Big(\delta^4(r)+\frac{a}{R_c}\delta^5(r)\Big),
\end{equation}

\noindent with

\begin{equation}
\delta^N(r)=\frac{1}{N}\Big[\Big(1-\frac{r}{R_c}-\frac{a}{R_c}\Big)^{-N}-\Big(1+\frac{r}{R_c}-\frac{a}{R_c}\Big)^{-N}\Big].
\end{equation}

\noindent In Eq. (\ref{pot_Kihara}), $z$ is the coordination number of the cell. These parameters, which depend on the structure of the clathrate and on the type of the cage (small or large), are given in Table \ref{cages}. The Kihara parameters $a$, $\sigma$ and $\epsilon$ for the molecule-water interactions, given in Table \ref{kihara}, have been taken from Diaz Pe\~na et al. (1982) for CO and from Parrish \& Prausnitz (1972a,b) for all the other molecules of interest.

Finally, the relative abundance $f_K$ of a guest molecule $K$ in a clathrate can be calculated with
respect to the whole set of species considered in the system as

\begin{equation}
\label{abondance} f_K=\frac{b_s y_{K,s}+b_\ell y_{K,\ell}}{b_s \sum_J{y_{J,s}}+b_\ell \sum_J{y_{J,\ell}}},
\end{equation}

\noindent where $b_s$ and $b_l$ are the number of small and large cages per unit cell respectively, for the considered clathrate structure.

Table \ref{trapping} gives the fraction of volatiles incorporated in H$_2$S, Xe and CH$_4$ dominated clathrates relative to their initial fraction available in the nebula gas ($F_{\rm i}$). Our calculations show that CO, N$_2$ and Ar are poorly trapped in clathrates. Indeed, their relative abundances in these structures are order-of-magnitudes lower than those found in the solar nebula. On the other hand, we note that substantial amounts of Xe and Kr are trapped in H$_2$S- and CH$_4$-dominated clathrates, respectively. About 18\% of available Xe is incorporated in H$_2$S-dominated clathrate prior the formation of Xe-dominated clathrate at lower temperature. Because the Kr/CH$_4$ ratio is larger in CH$_4$-dominated clathrate than in the solar nebula ($F_{\rm Kr}$ $>$ 1), we infer that all Kr is incorporated in this clathrate, thus preventing the formation of its pure condensate at lower temperature.

\section{Implications for the origin of Titan's atmosphere}
\label{sec:Titan}
\subsection{Partial devolatilization of Titan's planetesimals in Saturn's subnebula}
\label{sec:devol}

In order to explain the formation of Titan, we assume that it was formed in a low surface density circumSaturnian accretion disk, resulting from the reduced gas inflow characterizing the last stages of the planet's growth and refer the reader to the work of AM07 for a full description of the thermodynamic structure of the subnebula. In particular, using time-dependent accretion disk models, AM07 have shown that the evolution of the Saturn's subnebula is divided in two phases. During the first phase, the subnebula is fed through its outer edge by gas and gas-coupled solids originating from the protoplanetary disk. When the solar nebula disappears, the subnebula enters in its second phase of evolution. The mass flux through the outer edge stops, and the subnebula expands outward due to viscosity. In order to yield Titan from the Saturn's accretion disk, we assume that solid material has been supplied essentially by direct transport of gas-coupled solids into the disk with the gas inflow during the first phase of the subnebula's evolution or by capture of heliocentrically orbiting solids as they pass through the disk (Canup \& Ward 2002,2006). AM07 have proposed that once embedded in the subnebula, solids originating from Saturn's feeding zone could have been altered if they encountered gas temperature and pressure conditions high enough to generate a loss of volatiles during their migration, and if they remained relatively porous, which seems a reasonable assumption. While more calculations, taking into account migration of captured planetesimals, as well as their thermal evolution and devolatilization in an evolving subnebula, are necessary to quantitatively test the scenario of AM07, we favor this mechanism to explain the carbon monoxide and argon deficiencies in the atmosphere of Titan. Indeed, as Figs. 1 and 2 show, if planetesimals ultimately accreted by Titan experienced intrinsic temperatures of $\sim$ 50 K during their migration in Saturn's subnebula, they are expected to release most of their CO, Ar and N$_2$. However, more volatile compounds than CH$_4$ form pure condensates when the clathration efficiency is poor in the feeding zone of Saturn, and the gap between the devolatilization temperatures of CH$_4$-dominated clathrate and these ices is much larger than in the case of full clathration ($\Delta T \simeq$ 30 K instead of 8 K). Hence the hypothesis of poor clathration efficiency requires a less vigorous thermal control of planetesimals during their migration within Saturn's subnebula. 

Note that, in our scenario of partial devolatilization, a higher sublimation temperature of planetesimals is excluded since it would imply the dissociation of methane clathrate from solids accreted by Titan, a result in conflict with the large abundance of methane in the satellite's atmosphere. On the other hand, since Kr and Xe are incorporated at higher temperatures than $\sim$ 50 K in clathrates produced in the nebula, they cannot be eliminated via this partial devolatilization mechanism only.

\subsection{Mechanisms of noble gas trapping}
\label{sec:trapping}

In this section, we discuss the processes of noble gas trapping which might have occurred either in the solar nebula gas phase before the formation of ices that were ultimately accreted by Titan, or at the satellite's surface or atmosphere during its post-accretion evolution. Each of these processes may explain the deficiencies of Kr and Xe observed in the atmosphere of Titan.

It has been thus proposed that the presence of H$_3^+$ ion in the outer solar nebula may induce the trapping of Ar, Xe and Kr in the form of stable complexes XH$_3^+$ (with X = Ar, Kr and Xe; Pauzat \& Ellinger 2007; Mousis et al. 2008). The efficiency of this trapping is ruled by the abundance profile of H$_3^+$, which is poorly constrained in the primordial nebula. Assuming that the gas of the solar nebula is ionized by cosmic rays at a rate of 1 $\times$ 10$^{-15}$ s$^{-1}$, which is considered as an upper limit, Mousis et al. (2008) calculated that the abundance of H$_3^+$ in the 3 to 30 AU mid-plane region of the nebula is about 1 $\times$ 10$^{-12}$ relative to H$_2$, at most. Such a low concentration of H$_3^+$ implies a poor noble gas sequestration since the solar abundances of xenon, krypton and argon hold between about 10 $\times$ 10$^{-10}$ and 10 $\times$ 10$^{-6}$, relative to H$_2$ (see e.g. table \ref{lodders}). On the other hand, there is evidence that the solar nebula has been exposed to additional sources of energetic particles. This is testified by the presence of radionuclides with half-lives less than 1 Myr in the calcium-aluminum-rich inclusions in meteorites (see e.g. McKeegan \& Davis 2003) which could be due to the irradiation of energetic particles from the young Sun (Lee et al. 1998; Gounelle et al. 2001,2006) or to the presence of a nearby supernova during the solar system formation (Busso et al. 2003; Tachibana \& Huss 2003). If this hypothesis is correct, one associated effect might be an augmentation of the cosmic ionization rate in the disk by several orders of magnitude. As a result, given the noble gases abundances in the solar nebula, Kr and Xe could be relatively easily trapped, in regions where H$_3^+$ is at least $\sim$ 1 $\times$ 10$^{-9}$ times the H$_2$. From these considerations, Mousis et al. (2008) suggested that the H$_3^+$ abundance in the 10--30 AU region of the solar nebula was indeed effectively large enough to make possible at least the efficient trapping of xenon and krypton, and limited trapping of argon. Therefore, once formed, these complexes would remain stable, even at low temperature, and their presence in the outer nebula gas phase could have as one implication the formation of Kr and Xe-poor bodies that are then delivered to Titan.

Alternatively, it has been shown that if large amounts of Kr and Xe were initially present in Titan's atmosphere, they could have been efficiently trapped as clathrates by crystalline water ice located on the satellite's surface (Osegovic \& Max 2005; Thomas et al. 2007,2008). Indeed, by considering several initial gas phase compositions, including different sets of noble gases abundances that may be representative of Titan's early atmosphere, Thomas et al. (2007,2008) showed that the trapping efficiency of clathrates is high enough to significantly decrease the atmospheric concentrations of Xe and, in a lesser extent, of Kr, irrespective of the initial gas phase composition, provided that these clathrates are abundant enough at the surface of Titan. In these conditions, Thomas et al. (2007) calculated that the total sink of Xe or Kr in clathrates would represent a layer at the surface of Titan whose equivalent thickness would not exceed $\sim$ 50 cm. In addition, from laboratory experiments, Jacovi \& Bar-Nun (2008) recently proposed that the noble gases of Titan could be removed by their trapping in its atmospheric haze. Hence, these two proposed mechanisms, i.e. trapping in clathrates located on the surface or in the atmospheric haze could act as sinks of Xe and Kr in the atmosphere of Titan.

\section{Discussion}
\label{sec:discussion}

Taking into account the constraints derived from the current composition of Titan's atmosphere, and from precise calculations of multiple guest trapping in clathrates formed in the solar nebula, we propose a formation scenario in which the satellite accreted from solids initially produced in the solar nebula. Based on the conclusions of AM07, we postulate that during their drift in Saturn's subnebula, and prior to their accretion by the forming Titan, these solids encountered gas temperature and pressure conditions high enough to generate a loss via sublimation of most of their carbon monoxide and argon, but low enough to retain the incorporated methane, ammonia, krypton and xenon. Interestingly enough, our model predicts that if all CO and Ar pure condensates where devolatilized from planetesimals ultimately accreted by Titan, small amounts of these species remained essentially trapped in CH$_4$-dominated clathrates, and subsequently in the satellite, with CO/CH$_4$ $\sim$ $1.7 \times 10^{-3}$ and Ar/CH$_4$ $\sim$ $6.4 \times 10^{-5}$ (see e.g. Table \ref{trapping}), in good agreement with the abundances of CO and $^{36}$Ar observed in Titan's atmosphere.

A plausible gas phase composition was adopted for the primordial nebula, but the conclusions derived from our model are not affected by the choice of other molecular ratios. Indeed, whatever the range of CO:CO$_2$:CH$_3$OH:CH$_4$ and N$_2$:NH$_3$ initial gas phase molecular ratios adopted in the solar nebula, the stability curves of CO, N$_2$ and Ar ices produced in this environment still remain at lower temperatures than that of CH$_4$ clathrate. Hence, independent of the initial gas phase composition, icy planetesimals migrating within Saturn's subnebula are still first depleted in carbon monoxide, nitrogen and argon when they experience progressive heating. Moreover, our Titan formation scenario is not influenced by the possible inward migration of Saturn (Alibert et al. 2005a) during its formation within the nebula since the composition of planetesimals remains almost constant independent of formation distance in the disk, provided that there was a homogeneous gas phase in the solar nebula (Marboeuf et al. 2008).

Because more volatile compounds than CH$_4$ form pure condensates when the clathration efficiency is assumed to be poor in the feeding zone of Saturn, and since the gap between the devolatilization temperatures of CH$_4$-dominated clathrate and these ices is much larger than in the case of full clathration, a less vigorous thermal control of planetesimals is required during their migration within Saturn's subnebula. This suggests that the efficiency of clathration may have been limited in the outer solar nebula.

In situ measurements by a future probe of Kr and Xe abundances in Saturn's atmosphere could constrain the origin of their deficiency in Titan. If the abundances of Kr and Xe are detected in solar proportions in Saturn, this would support the hypothesis of an efficient sequestration of these species by H$_3^+$ in the primitive nebula gas phase prior the formation of solids. Note that the H$_3^+$ abundance is expected to increase with the growing heliocentric distance in the primitive nebula (Mousis et al. 2008). Hence, independent of Saturn's case, the greater than solar abundances of these noble gases measured by the Galileo probe in Jupiter's atmosphere (Owen et al. 1999) can be explained by the low H$_3^+$ concentration at the time of formation and location of its building blocks (Mousis et al. 2008). On the other hand, if the abundances of these noble gases are measured to be greater than solar in Saturn, this would favor the hypothesis of a Kr and Xe sequestration that occurred at a later epoch than that of the planetesimals formation in the giant planet's feeding zone. In this case, the future sampling of the material constituting the dunes of Titan, which is likely to derive from the stratospheric photolysis products (Jacovi \& Bar-Nun 2008), would give information on the potential trapping of noble gases by haze. Similarly, the sampling of the icy surface of Titan would allow the amount of noble gases possibly incorporated in clathrates to be quantified.

Because the temperature and pressure conditions increase within the subnebula at diminishing distances to Saturn (AM07), the planetesimals accreted by regular satellites interior to the orbit of Titan should have experienced the same or more devolatilization process during their migration. We predict that these satellites are depleted in primordial carbon monoxide and nitrogen, similar to Titan. Note that the INMS instrument aboard the Cassini spacecraft has detected 91 $\pm$ 3\% H$_2$O, 4 $\pm$ 1\% N$_2$ or CO\footnote{The resolution of the INMS instrument is too low to bring firm constraints on the nature of the observed species (Waite et al. 2006).}, 3.2 $\pm$ 0.6 \% CO$_2$, 1.6 $\pm$ 0.4 \% CH$_4$ and other minor compounds in the plumes of Enceladus (Waite et al. 2006; Hansen et al. 2006). The detection of CO or N$_2$ in the plumes may appear in conflict with our predictions but it has been proposed that N$_2$ could be the result of thermal processing of primordial NH$_3$ in the hot interior of Enceladus (Matson et al. 2007) or produced from the irradiation of solid ammonia on the surface of the satellite (Zheng et al. 2008). Moreover, it has been experimentally shown that UV irradiation of carbon grains embedded in water ice or of CH$_4$-containing icy mixtures could lead to the formation of CO and CO$_2$ molecules at low temperature (Mennella et al. 2006; Baratta et al. 2003). Therefore, all these processes suggest that the observed N$_2$ or CO may not be primordial in the plumes of Enceladus, in agreement with our Titan's formation scenario. A key measurement in Enceladus would be then the detection of argon since our scenario also predicts a strong depletion of this noble gas in satellites interior to Titan. More constraints on the composition of Enceladus will be placed by the extended Cassini mission, through further flybys closer to and within its active geysers.

\acknowledgements
This work was supported in part by the French Centre National d'Etudes Spatiales. Supports from the PID program ``Origines des Plan{\`e}tes et de la Vie" of the CNRS, and the {\it Cassini} project, are also gratefully acknowledged.

\clearpage

\clearpage

\clearpage

\begin{table}
\caption[]{Gas phase abundances in the solar nebula.}
\begin{center}
\begin{tabular}{lclc}
\hline
\hline
\noalign{\smallskip}
Species X &  (X/H$_2$)  & species X  & (X/H$_2$) \\	
\noalign{\smallskip}
\hline
\hline
\noalign{\smallskip}
O		& $1.16 \times 10^{-3}$		& N$_2$		& $4.05 \times 10^{-5}$ \\
C		& $5.82 \times 10^{-4}$ 		& NH$_3$   	& $4.05 \times 10^{-5}$ \\
N   		& $1.60 \times 10^{-4}$    		& CO      		& $2.21 \times 10^{-4}$ \\
S        	& $3.66 \times 10^{-5}$		& CO$_2$  	& $3.16 \times 10^{-5}$ \\
Ar       	& $8.43 \times 10^{-6}$		& CH$_3$OH  	& $6.31 \times 10^{-6}$ \\ 	
Kr       	& $4.54 \times 10^{-9}$		& CH$_4$  	& $3.16 \times 10^{-6}$ \\ 
Xe       	& $4.44 \times 10^{-10}$		& H$_2$S    	& $1.83 \times 10^{-5}$ \\
H$_2$O  	& $4.43 \times 10^{-4}$		&							     \\
\hline
\end{tabular}
\end{center}
\tablecomments{Elemental abundances derive from Lodders (2003). Molecular abundances result from the distribution of elements between refractory and volatile components (see text).}
\label{lodders}
\end{table}

\clearpage

\begin{table}[h]
\centering \caption{Parameters for the cavities.}
\begin{tabular}{lcccc}
\hline \hline
Clathrate structure & \multicolumn{2}{c}{I} & \multicolumn{2}{c}{II} \\
\hline
Cavity type     	& small     	& large     		& small     		& large \\
$R_c$ (\AA)     	& 3.975     & 4.300     	& 3.910     	& 4.730 \\
$b$             	& 2         	& 6         		& 16       		& 8     \\
$z$             	& 20        	& 24        		& 20        		& 28    \\
\hline
\end{tabular}\\
\tablecomments{$R_c$ is the radius of the cavity (values taken from Parrish \& Prausnitz 1972a,b). $b$ represents the number of small ($b_s$) or large ($b_\ell$) cages per unit cell for a given structure of clathrate (I or II), $z$ is the coordination number in a cavity.\\}
\label{cages}
\end{table}

\clearpage

\begin{table}[h]
\centering \caption{Parameters for the Kihara potential.}
\begin{tabular}{clccc}
\hline \hline
Ref     & Molecule   & $\sigma$(\AA)& $ \epsilon/k_B$(K)& $a$(\AA) \\
\hline
(a)		& H$_2$S		& 3.1558		& 205.85		& 0.36 \\
(a)		& CO$_2$	& 2.9681		& 169.09		& 0.36 \\
(a)        	& CH$_4$     	& 3.2398     	& 153.17     	& 0.300 \\
(a)        	& N$_2$       	& 3.2199     	& 127.95     	& 0.350 \\
(a)        	& Xe              	& 3.1906     	& 201.34     	& 0.280 \\
(a)        	& Ar              	& 2.9434     	& 170.50     	& 0.184 \\
(a)        	& Kr             	& 2.9739     	& 198.34     	& 0.230 \\
(b)       	& CO            	& 3.101      	& 134.95     	& 0.284 \\
\hline
\end{tabular}\\
\tablecomments{$\sigma$ is the Lennard-Jones diameter, $\epsilon$ is the depth of the potential well, and $a$ is the radius of the impenetrable core. These parameters derive from the works of (a) Parrish \& Prausnitz (1972a,b) and (b) Diaz Pe\~{n}a et al. (1982).\\}
\label{kihara}
\end{table}

\clearpage

\begin{table}[h]
\centering \caption{Abundance of volatile i in clathrate relative to initial abundance in the nebula.}
\begin{tabular}{lccc}
\hline \hline
Clathrate				& Species			& $f_{\rm i}$				& $F_{\rm i}$		\\
\hline
H$_2$S-dominated  	& CO$_2$		& $3.12 \times 10^{-6}$	& $1.81 \times 10^{-6}$	\\
					& Xe				& $4.36 \times 10^{-6}$	& 0.18				\\
					& CH$_4$      		& $5.45 \times 10^{-6}$     & $3.16 \times 10^{-5}$	\\
					& CO			& $2.13 \times 10^{-7}$	&  $1.76 \times 10^{-8}$	\\
					& Kr				& $3.57 \times 10^{-9}$	& $1.43 \times 10^{-5}$	\\
					& Ar				& $8.74 \times 10^{-9}$	& $1.90 \times 10^{-8}$	\\
					& N$_2$			& $4.04 \times 10^{-7}$	& $1.83 \times 10^{-7}$	\\
\hline
Xe-dominated 			& CH$_4$      		& $7.28 \times 10^{-2}$	& $1.02 \times 10^{-5}$	\\
					& CO			& $5.18 \times 10^{-5}$	& $1.04 \times 10^{-10}$	\\
					& Kr				& $7.82 \times 10^{-6}$	& $7.65 \times 10^{-7}$	\\
					& Ar				& $1.08 \times 10^{-6}$	& $5.69 \times 10^{-11}$	\\
					& N$_2$			& $6.68 \times 10^{-4}$	& $7.32 \times 10^{-9}$	\\
\hline
CH$_4$-dominated		& CO			& $1.74 \times 10^{-3}$	& $2.49 \times 10^{-5}$	 \\
					& Kr				& $1.67 \times 10^{-3}$	& 1.16			\\
					& Ar				& $6.43 \times 10^{-5}$	& $2.41 \times 10^{-5}$	\\
					& N$_2$			& $3.94 \times 10^{-3}$	& $3.07 \times 10^{-4}$	\\
\hline
\end{tabular}
\tablecomments{$f_{\rm i}$ is defined as the molar ratio of i to X in X-dominated clathrate and $F_{\rm i}$ is the ratio of $f_{\rm i}$ to the initial ratio (i to X) in the solar nebula. Calculations are performed at temperature and pressure conditions given by the intersection of the cooling curve at 9.5 AU and the stability curves of the considered clathrates (see Figs. \ref{cooling} and \ref{cooling2}). Only the species that are not yet condensed or trapped prior the epoch of clathrate formation are considered in our calculations.\\}
\label{trapping}
\end{table}

\end{document}